\documentclass[12pt]{article}

\usepackage{amsmath}
\usepackage{graphicx,psfrag,epsf}
\usepackage{enumerate}
\usepackage{psfrag,epsf,tikz}
\usepackage{url} 
\usepackage{amsmath,amssymb,amsfonts,amsthm,mathtools,mathrsfs}

\usepackage{algorithm}
\usepackage{algpseudocode}
\usepackage{booktabs}
\usepackage{bm,bbm}
\usepackage{color}
\usepackage{subfigure}
\usepackage{algorithm,algpseudocode}
\usepackage[symbol]{footmisc}
\usepackage{scalefnt}
\usepackage{authblk} 
\usepackage{multirow,centernot}
\usepackage[colorlinks=true, citecolor=blue, urlcolor=blue]{hyperref}
\usepackage{natbib}
\usepackage{dsfont}
\usepackage[title]{appendix}
\usepackage[left=1in,top=1.1in,right=0.5in,bottom=1in]{geometry}

\theoremstyle{definition}

\newtheorem{theorem}{Theorem}[section]

\newtheorem{proposition}[theorem]{Proposition}
\newtheorem{remark}[theorem]{Remark}

\makeatletter
\def\@seccntformat#1{\@ifundefined{#1@cntformat}%
	{\csname the#1\endcsname\quad}
	{\csname #1@cntformat\endcsname}
}
\makeatother

\newif\ifShowComments
\ShowCommentstrue
\def\strutdepth{\dp\strutbox}
\def\druk#1{\strut\vadjust{\kern-\strutdepth
        {\vtop to \strutdepth{%
                \baselineskip\strutdepth\vss
                        \llap{\hbox{#1}\quad}\null}}}}

\title{\bf Two Tunable Gini-Type Measures with U-Statistic Estimation: Theory, Simulation, and an Empirical Application to GDP per Capita in the Americas}

\author{
\text{Roberto Vila}$^{1}$\thanks{Corresponding author: Roberto Vila, email: {rovig161@gmail.com}}
\,\, and
\text{Helton Saulo}$^{1,2}$
\\
{\small $^{1}$ Department of Statistics, University of Brasilia, Brasilia, Brazil}\\
{\small $^{2}$ Department of Economics, Federal University of Pelotas, Pelotas, Brazil}\\
}
\begin{document}
	\maketitle 	
%
	\begin{abstract}
We introduce two families of inequality measures, $G_p$ and $H_q$, that converge to the classical Gini coefficient as $p,q\to\infty$. The tuning parameters $p>1$ and $q>0$ regulate the influence of disparities between observations. For each index we derive closed-form $U$-statistic plug-in estimators and establish strong consistency and asymptotic normality under mild moment conditions. 
A Monte Carlo study assesses finite-sample behavior across $(n,p,q)$, and an empirical illustration with GDP per capita in the Americas shows how the tuning parameters influence the measure of inequality.
\end{abstract}

%

\section{Introduction}
\label{sec:introduction}

Understanding and quantifying inequality in distributions of economic or social variables is a central concern across multiple disciplines, from economics \cite{YAO01101999} and environment studies \cite{SUN2010601} to health sciences \citep{Kharazmi2023} and ecology \citep{Damgaard2000}.  Among the array of summary metrics, the Gini coefficient \citep{Gini1936}, the normalized average absolute difference between two randomly chosen observations, has become the benchmark due to its clear interpretation and straightforward estimation \citep{Deltas2003}.  It is routinely employed by institutions such as the World Bank to track income and wealth disparities worldwide \citep{Baydil2025}.

Despite its widespread use, the classical Gini coefficient treats all pairwise gaps equally, which can mask nuanced patterns of dispersion. In this paper, we introduce two novel measures of inequality, denoted \(G_p\) and \(H_q\), that admit a tunable parameter to control sensitivity to tail‐inequality. The index \(G_p\), based on a logarithmic kernel as function of \(p>1\), and the index \(H_q\), derived from generalized sums as function of \(q>0\), both converge to the classical Gini as \(p,q\to\infty\).  We propose $U$-statistic-based plug-in estimators for \(G_p\) and \(H_q\) and establish their strong consistency and asymptotic normality under mild moment conditions using standard results \citep{Hoeffding1948}.

The remainder of the paper is organized as follows.  Section~\ref{Income inequality measures} defines the new measures.  Section~\ref{Income inequality estimators} presents the corresponding sample estimators.  In Section~\ref{Strong consistency}, we prove strong consistency, and Section~\ref{Asymptotic distribution} derives the asymptotic distributions.  Section~\ref{Simulation study} reports Monte Carlo results on finite-sample performance, and Section~\ref{sec:application} illustrates an empirical application to GDP per capita data.  Finally, in Section~\ref{Concluding remarks}, we provide some concluding remarks.

\section{Income inequality measures}\label{Income inequality measures}

Let $X_1$ and $X_2$ be two independent copies of a non-negative random variable $X$ with positive {\color{black} and finite} mean $\mu = \mathbb{E}(X)$. We define the following income inequality measures for $X$:
\begin{align}\label{Gp}
	G_{p}
	\equiv
	G_{p}(X)
	\equiv
		G_{p}(X_1,X_2)
	=
	\dfrac{\displaystyle
		{\mathbb{E}\left[\log\left(1+p^{X_2-X_1}\right)
		+
		\log\left(1+p^{X_1-X_2}\right)\right]
-
2\log(2)
}}{\displaystyle 2\log(p)\mu},
	\quad p>1,
\end{align}
and
\begin{align}\label{Hq}
	H_{q}
	\equiv
	H_{q}(X)
	\equiv
	H_{q}(X_1,X_2)
	=
	\dfrac{\displaystyle
		{\mathbb{E}\left[
		\left({X_1^{q}+X_2^{q}\over 2}\right)^{1/q}
		-
		\left({X_1^{-q}+X_2^{-q}\over 2}\right)^{-1/q}
			\right]
	}}{\displaystyle 2\mu},
	\quad q>0.
\end{align}

It is worth noting that as $p$ and $q$ both increase, the above measures approach the  classical definition of the Gini coefficient \citep{Gini1936}, denoted by $G$ (see Proposition \ref{Limit behavior}).

{\color{black}The classical Gini coefficient assigns equal weight to every pairwise difference $|X_1-X_2|$. The measure $G_p$ replaces the raw difference with the scaled kernel
\[
T(p) \equiv \frac{\log(1+p^{X_2-X_1})+\log(1+p^{X_1-X_2})-2\log(2)}{\log(p)},
\]
so that $G_p = \mathbb{E}[T(p)]/(2\mu)$. For finite $p>1$, $T(p)$ is an increasing and concave function of $|X_1-X_2|$ that grows more slowly than $|X_1-X_2|$ itself: pairs with very large differences contribute proportionally less to $G_p$ than they do to the Gini, while pairs with moderate differences receive relatively more weight. As $p\to\infty$, $T(p)\to|X_1-X_2|$ and the Gini is recovered. The parameter $p$ therefore controls the degree of inequality aversion in a manner analogous to the parameter in the Atkinson index. An analogous interpretation holds for $H_q$ via the power-mean difference kernel $K(q) \equiv M_q(X_1,X_2)-M_{-q}(X_1,X_2)$, where $M_r$ denotes the $r$-th power mean.

It is relevant to note that these measures are designed not to replace the Gini but to complement it through a sensitivity analysis. By examining the profile $p\mapsto\widehat{G}_p$ (or $q\mapsto\widehat{H}_q$) across a range of values, the analyst gains insight into the distributional structure of inequality: a flat profile indicates that inequality is spread uniformly across pairwise gaps and the Gini is a robust summary; a steeply increasing profile signals that extreme disparities are the dominant driver of inequality.
}

\begin{remark}\label{rem-initial}
The following relationships are noteworthy:
\begin{align*}
\min\{X_1,X_2\}
\leqslant
\log\left[\left(\dfrac{p^{X_1}+p^{X_2}}{2}\right)^{1/\log(p)}\right]
 ,
 \left({X_1^{q}+X_2^{q}\over 2}\right)^{1/q}
 \leqslant
 \max\{X_1,X_2\}
\end{align*}
and
\begin{align*}
	-\max\{X_1,X_2\}
	\leqslant
	\log\left[\left(\dfrac{p^{-X_1}+p^{-X_2}}{2}\right)^{1/\log(p)}\right]
	\leqslant
	-
	\min\{X_1,X_2\}.
\end{align*}
Hence,
\begin{multline*}
\left\vert
		\dfrac{\displaystyle
		{\log\left(1+p^{X_2-X_1}\right)
			+
			\log\left(1+p^{X_1-X_2}\right)
			-
			2\log(2)
	}}{\displaystyle \log(p)}
\right\vert
\\[0,2cm]
=
\left\vert
\log\left[\left(\dfrac{p^{X_1}+p^{X_2}}{2}\right)^{1/\log(p)}\right]
+
	\log\left[\left(\dfrac{p^{-X_1}+p^{-X_2}}{2}\right)^{1/\log(p)}\right]
\right\vert
\\[0,2cm]
\leqslant
 \max\{X_1,X_2\}
 -
 	\min\{X_1,X_2\}
 	\leqslant
\max\{X_1,X_2\}
\end{multline*}
and
\begin{multline*}
	\left\vert
	\left({X_1^{q}+X_2^{q}\over 2}\right)^{1/q}
	-
	\left({X_1^{-q}+X_2^{-q}\over 2}\right)^{-1/q}
	\right\vert
	\leqslant
	\left({X_1^{q}+X_2^{q}\over 2}\right)^{1/q}
+
\left({X_1^{-q}+X_2^{-q}\over 2}\right)^{-1/q}
\leqslant
2\max\{X_1,X_2\}.
\end{multline*}
\end{remark}

{\color{black}
	{\color{black}
In what follows, Propositions \ref{Monotonicity}--\ref{pro-eq-2} present results and characterizations of the income inequality measures $G_p$ and $H_q$.
}
	\begin{proposition}[Existence]\label{prop-1}
{\color{black}		
The index $G_p$ exists for all $p>1$, and the index $H_q$ exists for all $q>0$.
}
\end{proposition}
\begin{proof}
{\color{black}
By Remark \ref{rem-initial}, the expectations
$\mathbb{E}[\log(1+p^{X_2-X_1})]$, 
$\mathbb{E}[\log(1+p^{X_1-X_2})]$, 
$\mathbb{E}[({(X_1^{q}+X_2^{q})}/{2})^{1/q}]$, 
$\mathbb{E}[({(X_1^{-q}+X_2^{-q})}/{2})^{-1/q}]$
exist. Hence, the indices $G_p$ for $p>1$ and $H_q$ for $q>0$ are well defined.
} 
\end{proof}

{\color{black}
The following result shows that the indices $G_p$ and $H_q$ exhibit increasing inequality aversion.
}
\begin{proposition}[Monotonicity in parameters $p$ and $q$]\label{Monotonicity}
The following {\color{black} statements} hold:  
\begin{enumerate}
\item  $G_p$ is increasing in $p$.
\item $H_q$ is increasing in $q$. 
\end{enumerate}
{\color{black} Moreover, if $\mathbb{P}(X_1 \neq X_2) > 0$, then both monotonicities are strict.}
\end{proposition}
{\color{black}
\begin{proof}
Setting $t = \log(p) > 0$, $\psi(x) = 2\log(\cosh(x/2))$ and $\phi(x,t)=\psi(tx)/{(2\mu t)}$, we can rewrite
\[
G_p(t) \equiv G_p
= \frac{\mathbb{E}[\psi(t(X_2 - X_1))]}{2\mu t}
=
\mathbb{E}[\phi(X_2-X_1,t)].
\]
Since $t = \log(p)$ is increasing in $p>1$, it suffices to show that $t\mapsto G_p(t)$ is increasing in $t$.

Indeed, 
let $t \in [a,b]\subset(0,\infty)$. Since $|\psi'(x)|= |\tanh(x/2)| \leqslant 1$ and $|\psi(x)| \leqslant C_0(1+|x|)$, we obtain
$
\left|{\partial \phi(x_2-x_1,t)}/{\partial t}\right|
\leqslant 
C(1+|x_2-x_1|)
\equiv
B(x_2-x_1),
$
for some constant $C=C(a,b,\mu)>0$. In particular, the bound is uniform in $t\in[a,b]$. Since $\mathbb{E}|X|<\infty$, it follows that $\mathbb{E}|X_2-X_1|<\infty$, and hence $\mathbb{E}[B(X_2-X_1)]<\infty$. Therefore, by the dominated convergence theorem,
\[
G_p'(t)
= 
\mathbb{E}\left[{\partial \phi(X_2-X_1,t)\over \partial t}\right]
=
\frac{\mathbb{E}\!\big[t(X_2 - X_1)\,\psi'(t(X_2 - X_1)) - \psi(t(X_2 - X_1))\big]}{2\mu t^2}.
\]
Since $\psi'(x) = \tanh(x/2)$, the numerator becomes
\[
\mathbb{E}\!\left[
t(X_2 - X_1) \tanh\!\left(\frac{t(X_2 - X_1)}{2}\right)
- 2\log\!\left(\cosh\!\left(\frac{t(X_2 - X_1)}{2}\right)\right)
\right].
\]
Using the inequality
$
x \tanh(x/2) \geqslant 2\log(\cosh(x/2)), \ x \in \mathbb{R},
$
with strict inequality for $x \neq 0$, and applying it with $x = t(X_2 - X_1)$, the integrand is nonnegative. Hence $G_p'(t) \geqslant 0$, so $G_p(t)$ is increasing in $t$. Moreover, if $\mathbb{P}(X_1 \neq X_2) > 0$, then the monotonicity is strict. This proves Item~1.

For Item~2, let $x_1,x_2>0$ and define the power mean
$
M_r(x_1,x_2) \equiv \left({(x_1^r + x_2^r)}/{2}\right)^{1/r}.
$
It is well known that, for fixed $x_1,x_2>0$, the function $r \mapsto M_r(x_1,x_2)$ is increasing in $r$, and strictly increasing if $x_1 \neq x_2$. Therefore,
$
q \mapsto
M_q(x_1,x_2) - M_{-q}(x_1,x_2)
$
is increasing in $q>0$, strictly so whenever $x_1 \neq x_2$.
Applying this pointwise with $(x_1,x_2) = (X_1,X_2)$ and taking expectations, we conclude that $H_q$ is increasing in $q>0$. Moreover, if $\mathbb{P}(X_1 \neq X_2) > 0$, the monotonicity is strict. This completes the proof of Item~2 and hence the proposition.
\end{proof}
}

\begin{proposition}[Limit behavior]\label{Limit behavior}
	The following hold:
	\begin{align*}
		\lim_{p\to\infty}	G_{p}
		=
		\lim_{q\to\infty}	H_{q}
		=
		{\mathbb{E}\left[\max\{X_1,X_2\}-\min\{X_1,X_2\}\right]\over 2\mu}
		=
		{\mathbb{E} \vert X_1-X_2\vert \over 2\mu}
		=
		G.
	\end{align*}
	Furthermore, $\lim_{p\to 1^+}	G_{p}
	=
	\lim_{q\to 0^+}	H_{q}
	=0.$
\end{proposition}
\begin{proof}
The proof is straightforward and thus omitted.
\end{proof}

\begin{proposition}[Symmetry]
	The measures $G_p$ and $H_q$ are invariant under symmetries of the population: $G_p(X_1,X_2)=G_p(X_2,X_1)$ and $H_q(X_1,X_2)=H_q(X_2,X_1)$.
\end{proposition}
\begin{proof}
	The result follows directly from the definitions of $G_p$ and $H_q$.
\end{proof}

\begin{proposition}[Normalization]
		The measures $G_p$ and $H_q$ are normalized to lie between 0 and 1; that is, $0\leqslant G_p\leqslant 1$ and $0\leqslant H_q\leqslant 1$.
\end{proposition}
\begin{proof}
The proof follows immediately from Remark \ref{rem-initial}.
\end{proof}

	\begin{proposition}\label{prop-invariance-1}
	The	measure $G_p$ satisfies the following properties:
	\begin{enumerate}
		\item  \textbf{Non ratio-scale invariance:}	 $G_p$ is not invariant under positive scalar multiplication of the variable $X$; that is, for any $b > 0$,	
		$$
		G_p(bX)=G_{{\color{black} p^b}}(X).
		$$

		\item \textbf{Lack of translation invariance:} 
		$G_p$ is affected by additive shifts in \( X \), meaning it is not translation invariant. More precisely, for any \( a > 0 \),
		\[
		G_p(a + X) = \left(\frac{\mu}{a + \mu}\right) G_p(X),
		\]
		where \( \mu \) denotes the mean of \( X \).
	\end{enumerate}
\end{proposition}
\begin{proof}
	The proof follows directly from the definition \eqref{Gp} of $G_p$ and is therefore omitted.
\end{proof}

\begin{proposition}\label{prop-invariance-2}
		The	measure $H_q$ satisfies the following properties:
\begin{enumerate}
	\item 
	\textbf{Ratio-scale invariance:}	 $H_q$ is invariant under positive scalar multiplication of the variable $X$; that is, for any $b > 0$,	
	$$
	H_q(bX)=H_q(X).
	$$
	\item \textbf{Lack of translation invariance:} 
	$H_q$ fails to be translation invariant, since for any \( a > 0 \),
	\begin{align*}
	H_q(a + X) \neq H_q(X).
	\end{align*}
\end{enumerate}
\end{proposition}
\begin{proof}
		Since the result follows directly from the definition \eqref{Hq} of $H_q$, the proof is omitted.
\end{proof}

{\color{black}
\begin{remark}[Comparison between $G_p$ and $H_q$]\label{rem-comparison}
The two measures differ in several important respects. First, $H_q$ is ratio-scale invariant, i.e., $H_q(bX)=H_q(X)$ for any $b>0$, a property shared by the classical Gini coefficient; in contrast, $G_p$ satisfies $G_p(bX)=G_{p^b}(X)$, so rescaling income is equivalent to changing the tuning parameter. In practice, $H_q$ is preferable when the measurement unit of income should not affect the inequality assessment, whereas $G_p$ may be appropriate when one wishes to study how the scale of income interacts with inequality sensitivity. Second, the logarithmic kernel of $G_p$ is more sensitive to extreme pairwise gaps than the power-mean kernel of $H_q$: for the same parameter value, $G_p$ tends to produce larger estimates (see Table~\ref{tab:estimates} in Section~\ref{sec:application}). Finally, both measures converge to the classical Gini as $p,q\to\infty$, but the rates of convergence depend on the underlying distribution in a non-comparable way, since $p$ and $q$ operate through different functional forms.
\end{remark}
}

\begin{proposition}\label{pro-eq-1}
	For every $p>1$ one can find a value $r_p> 0$
	satisfying 
	\begin{align*}
		G_p(X)=G(r_p+X),
	\end{align*}
	where $G$ is the classical Gini coefficient \citep{Gini1936}. {\color{black} Moreover, if $X$ is non-degenerate, the value $r_p$ is unique.
	}
\end{proposition}
\begin{proof}
	Consider the notations  $s\equiv G_p(X)$ and the function  $f(r)\equiv G(r+X)$, for $r\geqslant 0$. {\color{black} Note that $f$ is non-increasing on $[0,\infty)$, and strictly decreasing unless $X$ is almost surely constant, in which case $G(r+X)=0$ for all $r \geqslant 0$. Moreover, since $f$} satisfies the boundary conditions
	$f(\infty)=0<s{\, \color{black} = \,} G_p(X)\leqslant G(X)=f(0^+)$,
	 the intermediate value Theorem  
{\color{black}	 
guarantees the existence of a point $r_p \in (0,\infty)$ satisfying $f(r_p)=s$, which is unique when $X$ is non-degenerate.
}	 
	 This concludes the proof.
\end{proof}

\begin{remark}
	Since the Gini coefficient $G$ is not translation invariant, by using Proposition \ref{pro-eq-1}, we have
	$
	G_p(X) = [{\mu}/({r_p + \mu})] G(X),
	$
	from which we get
	\begin{align*}
	r_p=\mu\left[{G(X)\over G_p(X) }-1\right]\geqslant 0.
	\end{align*}
	It is clear that when $p\to\infty$, we get $r_p=0$.
\end{remark}

\begin{proposition}\label{pro-eq-2}
	For every $q>0$ one can find a value ${\color{black}t_q}> 0$
	satisfying 
	\begin{align*}
		H_q(X)=G({\color{black}t_q}+X),
	\end{align*}
	where $G$ is the classical Gini coefficient. {\color{black} Moreover, if $X$ is non-degenerate, the value $t_q$ is unique.
	}
\end{proposition}
\begin{proof}
	The proof follows the same steps as Proposition \ref{pro-eq-1} and is therefore omitted.
\end{proof}

\begin{remark}
	Propositions \ref{pro-eq-1} and \ref{pro-eq-2} provide a meaningful interpretation of the indices $G_p$ and $H_q$
	within the classical Gini framework. Specifically, these measures are equivalent to evaluating the standard Gini coefficient on the adjusted variables 
$r_p+X$ and ${\color{black}t_q}+X$, where the adjustment parameters 
$r_p$ and ${\color{black}t_q}$ quantify the extent of inequality aversion embedded in the indices
\end{remark}
}

\section{Income inequality estimators} \label{Income inequality estimators}

The income inequality estimators of indices $G_{p}$ and $H_{q}$, defined in Section \ref{Income inequality measures}, are defined as follows (for $p>1$ and $q>0$):
\begin{align}\label{estimator}
	\widehat{G_{p}}
	=
	\dfrac{\displaystyle
			\sum_{1\leqslant i<j\leqslant n}
		\left[
		{\log\left(1+p^{X_{j}-X_{i}}\right)
			+
			\log\left(1+p^{X_{i}-X_{j}}\right)
			-
			2\log(2)
		}
		\right]
	}{\displaystyle {\color{black} (n-1)}\log(p)\sum_{i=1}^{n}X_i}
\end{align}
and
\begin{align}\label{estimator-1}
	\widehat{H_{q}}
	=
	\dfrac{\displaystyle
		\sum_{1\leqslant i<j\leqslant n}
		\left[
		\left({X_{i}^{q}+X_{j}^{q}\over 2}\right)^{1/q}
		-
		\left({X_{i}^{-q}+X_{j}^{-q}\over 2}\right)^{-1/q}
		\right]
	}{\displaystyle {\color{black} (n-1)}\sum_{i=1}^{n}X_i},
\end{align}
respectively,
where $X_1,\ldots, X_n$ are iid observations from the  population $X$.

\begin{remark}
	Setting $p\to\infty$ and $q\to\infty$ in \eqref{estimator} and \eqref{estimator-1}, respectively, we get the 
	{\color{black}\cite{Deltas2003} modified estimator of the Gini coefficient},
	denoted by $\widehat{G}$,
	\begin{align*}
		\lim_{p\to\infty}	\widehat{G_{p}}
		=
		\lim_{q\to\infty}	\widehat{H_{q}}
		&=
\dfrac{\displaystyle
	\sum_{1\leqslant i<j\leqslant n}
	\left[
	\max\{X_{i},X_{j}\}-\min\{X_{i},X_{j}\}
	\right]
}{\displaystyle {\color{black} (n-1)}\sum_{i=1}^{n}X_i}
\\[0,2cm]
		&=
		\dfrac{\displaystyle\sum_{1\leqslant i<j\leqslant n} \vert X_i-X_j\vert}{\displaystyle {\color{black} (n-1)}\sum_{i=1}^{n}X_i}
		=
		\widehat{G}.
	\end{align*}
\end{remark}

\begin{figure}[htbp]
    \centering
    \includegraphics[width=0.75\textwidth]{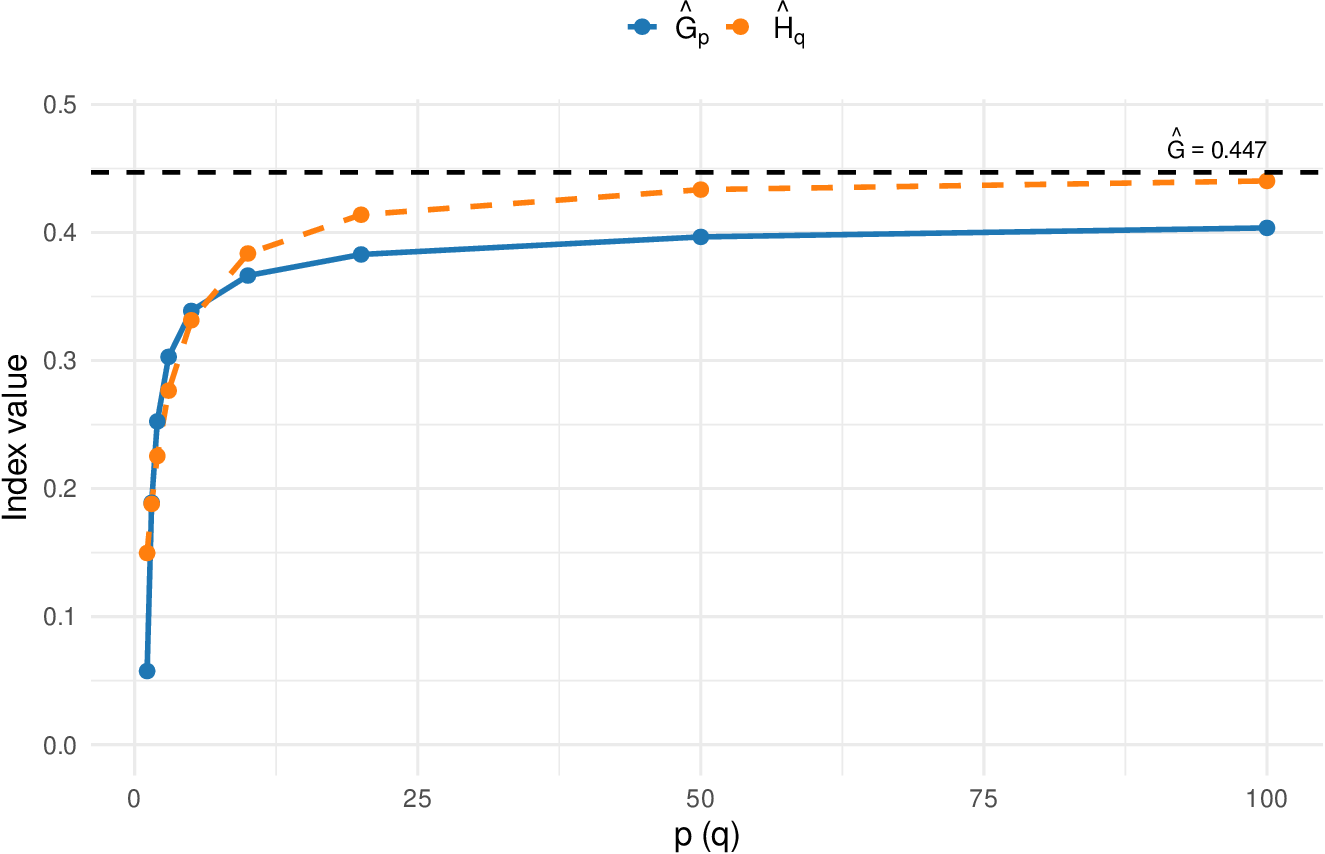}
    \caption{Behavior of the estimators $\widehat{G}_p$ and $\widehat{H}_q$ defined in equations~(3) and~(4), respectively, as functions of $p$ and $q$. The dashed line represents the classical Gini estimator $\widehat{G}$. Estimates were computed using a sample of size $n = 50$ drawn from a Gamma distribution with shape 1.5 and scale 2.5.}
    \label{fig:ineq_estimators_sim}
\end{figure}

Figure~\ref{fig:ineq_estimators_sim} illustrates the behavior of the estimators $\widehat{G}_p$ and $\widehat{H}_q$ defined in \eqref{estimator} and \eqref{estimator-1}, respectively, for increasing values of $p$ and $q$, based on a gamma sample. {\color{black} From this figure, we observe that both $\widehat{G}_p$ and $\widehat{H}_q$ are monotone increasing and converge to the classical Gini estimator $\widehat{G}$ as their respective parameters grow, consistently with the theoretical limit established in Proposition~\ref{Limit behavior}. It is worth noting that $p$ and $q$ are not on the same scale and operate through different functional forms, so no direct comparison of their convergence rates is meaningful.}

\begin{figure}[!ht]
  \centering
  \includegraphics[width=0.6\textwidth]{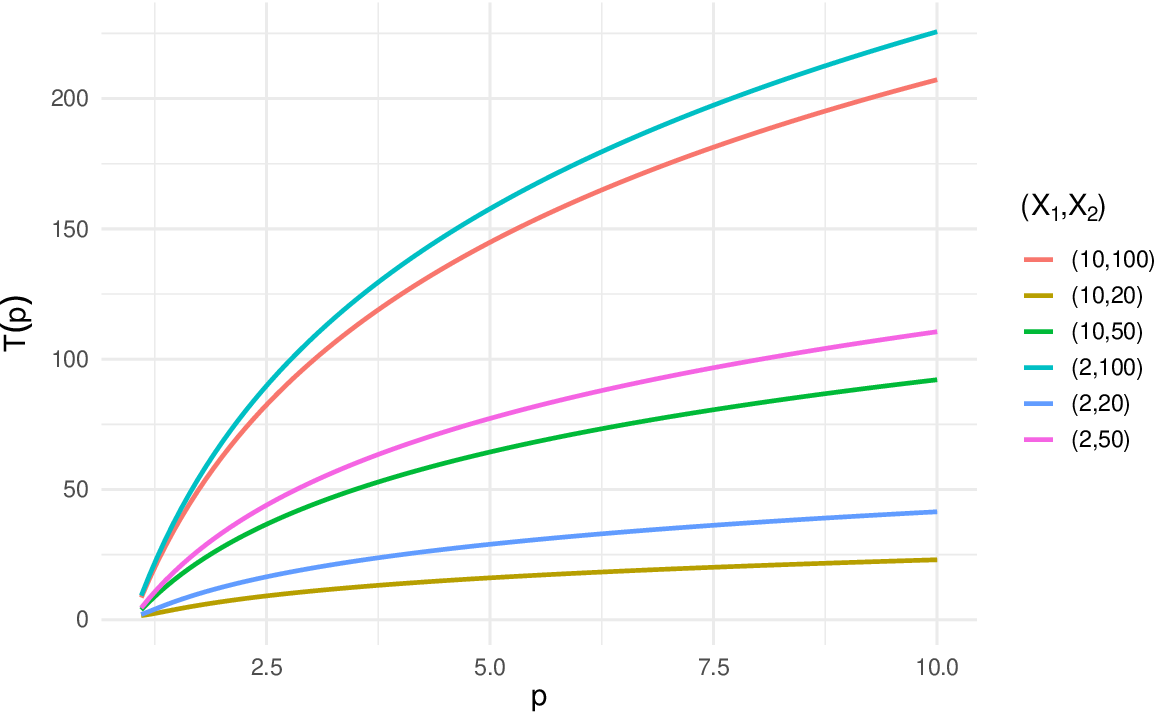}
  \caption{{\color{black} Curves of
    $T(p)=[\log(1+p^{X_2 - X_1})+\log(1+p^{X_1 - X_2})-2\log 2]/\log p$
    for various pairs $(X_1,X_2)$.}}
  \label{fig:Tp_overlay}
\end{figure}

\begin{figure}[!ht]
  \centering
  \includegraphics[width=0.6\textwidth]{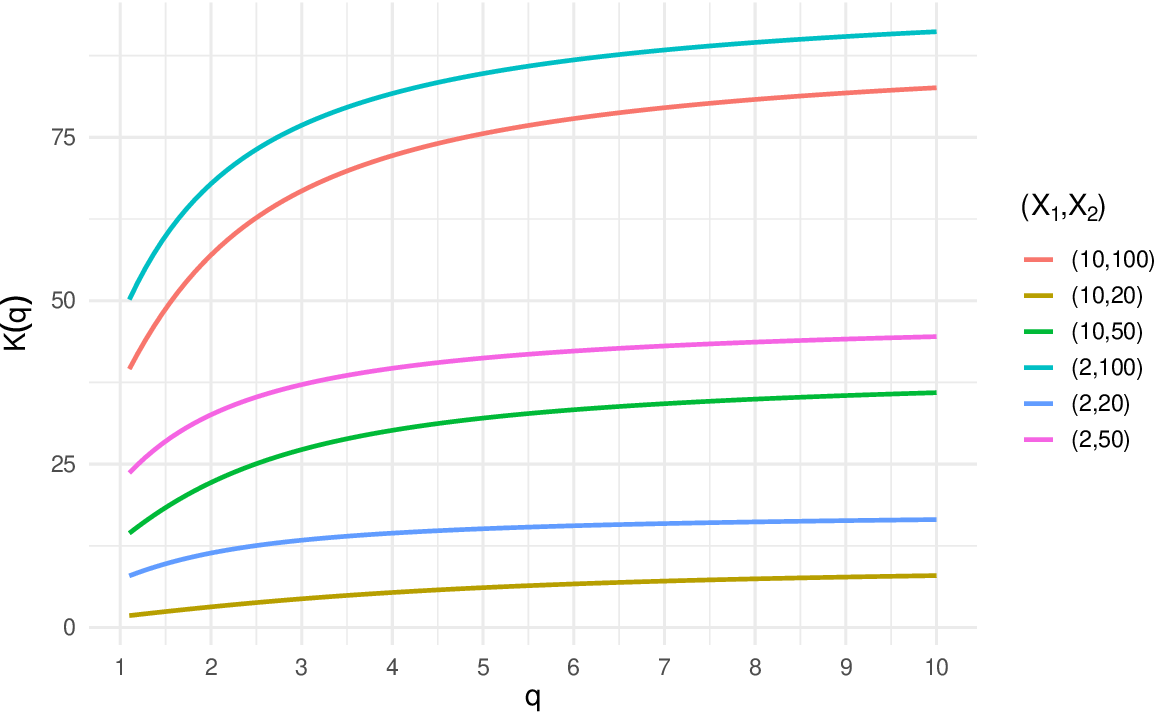}
  \caption{Curves of
    $K(q)
      =\bigl(\tfrac{X_1^q + X_2^q}{2}\bigr)^{1/q}
       -\bigl(\tfrac{X_1^{-q}+X_2^{-q}}{2}\bigr)^{-1/q}$
    for several pairs \((X_1,X_2)\).}
  \label{fig:Kq_overlay}
\end{figure}

{\color{black} In order to understand the role of $p$ and $q$ on the computation of $\widehat G_p$ and $\widehat H_q$, we plot the values of $p$ and $q$ against}
\begin{align*}
T(p)
\equiv
\frac{\log\bigl(1+p^{X_2 - X_1}\bigr)\;+\;\log\bigl(1+p^{X_1 - X_2}\bigr) - 2\log(2)}{\log(p)}
\quad
\text{and}
\quad
K(q)
\equiv
\bigl(\tfrac{X_1^q + X_2^q}{2}\bigr)^{1/q}
       -\bigl(\tfrac{X_1^{-q}+X_2^{-q}}{2}\bigr)^{-1/q},
\end{align*}
{\color{black} respectively; see Figures~\ref{fig:Tp_overlay} and~\ref{fig:Kq_overlay}. With this scaling, $G_p = \mathbb{E}[T(p)]/(2\mu)$, so the curves of $T(p)$ are directly relevant to the behavior of $G_p$. As $p$ increases, pairs with larger differences yield progressively steeper curves for $T(p)$, meaning that $p$ assigns greater weight to extreme pairwise disparities; for pairs with equal values ($X_1=X_2$), $T(p)=0$ for all $p$. The index $G_p$ for a given $p$ is obtained by integrating $T(p)$ against the joint distribution of $(X_1,X_2)$, so it reflects a distributional average of these curves. An analogous interpretation holds for $K(q)$: as $q$ increases, the power-mean difference $K(q)$ becomes more sensitive to the gap between the largest and smallest values in each pair, and $H_q = \mathbb{E}[K(q)]/(2\mu)$ consequently assigns greater weight to extreme disparities.}

\section{Strong consistency}\label{Strong consistency}

	Note that $	\widehat{G_{p}}$ in \eqref{estimator} can be written as
	\begin{align*}
	\widehat{G_{p}}
	=
	\dfrac{	U_{n}}{2\overline{X}},
	\end{align*}
where $\overline{X}=\sum_{i=1}^{n}X_i/n$ is the sample mean,
\begin{align}\label{U-n-def}
	U_{n}
		\equiv
		\displaystyle
		{\displaystyle\binom{n}{2}^{-1}}
		\sum_{1\leqslant i<j\leqslant n}
		g(X_i,X_j)
\end{align}
is a $U$-statistic \citep{Hoeffding1948} and
	\begin{align}\label{g-def}
	g(X_i,X_j)
	\equiv
	{
			{\log\left(1+p^{X_{j}-X_{i}}\right)
		+
		\log\left(1+p^{X_{i}-X_{j}}\right)
		-
		2\log(2)
		\over
		\log(p)}
	}.
	\end{align}

	If $\mathbb{E}\left\vert g(X_1,X_2)\right\vert<\infty$, then,  by strong law of large numbers for $U$-Statistics  \citep{Lee1990,Henze2024},
	\begin{align*}
	U_{n}\stackrel{\text{a.s.}}{\longrightarrow}
\mathbb{E}\left[ g(X_1,X_2)\right],
\quad 
 \text{as} \quad n\to\infty,
	\end{align*}
	with $\stackrel{\text{a.s.}}{\longrightarrow}$ meaning almost sure convergence. 
	Since $\overline{X}\stackrel{\text{a.s.}}{\longrightarrow}\mu$, it follows from the properties of almost sure convergence that
		\begin{align*}
		\widehat{G_{p}}
		=
		\dfrac{	U_{n}}{2\overline{X}}
		\stackrel{\text{a.s.}}{\longrightarrow}
\dfrac{\displaystyle
	{\mathbb{E}\left[ g(X_1,X_2)\right]
}}{\displaystyle 2\mu}
=
G_{p},
\end{align*}
where $G_{p}$ is as defined in \eqref{Gp}.

\smallskip
{\color{black}
On the other hand, observe that
\begin{align}\label{def-h-new}
	\widehat{H_{q}}
	&=
	\frac{V_{n}}{2\overline{X}}, 
	\qquad 
	V_{n}
	\equiv 
	\binom{n}{2}^{-1}
	\sum_{1\leqslant i<j\leqslant n}
	h(X_i,X_j),  \nonumber
	\\[0,2cm]
	h(X_1,X_2)
	&\equiv
	\left(\frac{X_{1}^{q}+X_{2}^{q}}{2}\right)^{1/q}
	-
	\left(\frac{X_{1}^{-q}+X_{2}^{-q}}{2}\right)^{-1/q}.
\end{align}
}

Under the condition $\mathbb{E}\big| h(X_1,X_2)\big|<\infty$, {\color{black}it follows from the strong law of large numbers that}, as $n\to\infty$,
\begin{align*}
	\widehat{H_{q}}
	\stackrel{\text{a.s.}}{\longrightarrow}
	\frac{\mathbb{E}\!\left[ h(X_1,X_2)\right]}{2\mu}
	=
	H_{q},
\end{align*}
where $H_{q}$ is {\color{black} defined} in \eqref{Hq}.

\begin{remark}
	From Remark \ref{rem-initial} we have $\left\vert g(X_1,X_2)\right\vert\leqslant \max\{X_1,X_2\}$ and $\left\vert h(X_1,X_2)\right\vert\leqslant 2\max\{X_1,X_2\}$. Then,  a sufficient condition for $\mathbb{E}\left\vert g(X_1,X_2)\right\vert<\infty$ and $\mathbb{E}\left\vert h(X_1,X_2)\right\vert<\infty$ is that
	$
	\mathbb{E}[\max\{X_1,X_2\}]<\infty.
	$
	But since $\max\{X_1,X_2\}\leqslant X_1+X_2$, it is sufficient that both random variables, $X_1$ and $X_2$, have finite expectations.
\end{remark}

\section{Asymptotic distribution}\label{Asymptotic distribution}


	It is well-known that \citep[Theorem 7.3 of][]{Hoeffding1948}, if $\mathbb{E}\left[g^2(X_1,X_2)\right]<\infty$ and
	$\mathbb{E}[X_1^2]<\infty$
	then,  as $n\to\infty$,
	\begin{align*}
	{\sqrt{n}\left\{
		\begin{pmatrix}
		U_n
		\\[0,2cm]
		\overline{X}
		\end{pmatrix}
	-
	\begin{pmatrix}
	\mathbb{E}\left[ g(X_1,X_2)\right]
	\\[0,2cm]
	\mu
	\end{pmatrix}
	\right\}}
	\stackrel{\mathscr{D}}{\longrightarrow}
	N_2\left(	
		\bf 0,
	\bf\Sigma
	\right),
	\end{align*}
	where 
	$\stackrel{\mathscr{D}}{\longrightarrow} $ denotes convergence in distribution, $U_n$ is the $U$-statistics defined in \eqref{U-n-def}, $g(X_1,X_2)$ is as in \eqref{g-def},  
		{\color{black} $\mathbf{0}=(0,0)^\top$} and $\mathbf{\Sigma}$ is the covariance matrix {\color{black} given by
		\begin{align}\label{covariance-matrix}
		\mathbf{\Sigma}
		=
		\begin{pmatrix}
			4\xi^{(1)}_g & 2\xi^{(1,2)}_g 
			\\[0,2cm]
			2\xi^{(1,2)}_g & \xi^{(2)}
		\end{pmatrix},
		\end{align}
		}
		whose elements are defined according to the following quantities:
\begin{align}\label{def-xi}
	\begin{array}{lll}
	&\xi^{(1)}_g
	\equiv
	\mathbb{E}_{X_1}\left\{\mathbb{E}_{X_2}^2[g(X_1,X_2)]\right\}
	-
	\mathbb{E}^2[g(X_1,X_2)],
	\\[0,2cm]
	&\xi^{(2)}
	\equiv
	\text{Var}(X_1),
	\\[0,2cm]
	&\xi^{(1,2)}_g
	\equiv
	\mathbb{E}_{X_1}\left\{X_1\mathbb{E}_{X_2}[g(X_1,X_2)]\right\}
	-
	\mathbb{E}[X_1] \mathbb{E}[g(X_1,X_2)].
		\end{array}
\end{align}
In the above, $\mathbb{E}_{X_2}[g(X_1,X_2)]$ indicates that the expectation is computed over the distribution of $X_2$, treating $X_1$ as fixed. 

For a given function $\vartheta$ with continuous first partial derivatives and a specific value of  $(\mathbb{E}\left[ g(X_1,X_2)\right],\mu)^\top$
for which ${\bf A}{\bf\Sigma}{\bf A}^{\top}>0$, {\color{black} where ${\bf\Sigma}$ is as given in \eqref{covariance-matrix} and
${\bf A}$ is a $1\times 2$ matrix defined by
\begin{align*}
	{\bf A}
	\equiv
	\left({\partial \vartheta(x,y)\over \partial x} \quad  {\partial \vartheta(x,y)\over  \partial y}\right)\Bigg\vert_{x=\mathbb{E}\left[ g(X_1,X_2)\right],y=\mu},
\end{align*}
}
the multivariate delta method provides
	\begin{align}\label{5-conv}
	{\sqrt{n}\left\{
		\vartheta
		\begin{pmatrix}
			U_n
			\\[0,2cm]
			\overline{X}
		\end{pmatrix}
		-
		\vartheta
		\begin{pmatrix}
			\mathbb{E}\left[ g(X_1,X_2)\right]
			\\[0,2cm]
			\mu
		\end{pmatrix}
		\right\}}
	\stackrel{\mathscr{D}}{\longrightarrow}
	N\left(	
	0
,
{\bf A}{\bf\Sigma}{\bf A}^{\top}
	\right).
\end{align}

\smallskip 
By taking $\vartheta(x,y)=x/(2y)$ in \eqref{5-conv},  we get
	\begin{align}
		\sqrt{n}(\widehat{G_{p}}-G_p)
		&=
		\sqrt{n}
		\left\{
		\dfrac{U_{n}}{2\overline{X}}
		-
		\dfrac{\mathbb{E}\left[ g(X_1,X_2)\right]
		}{\displaystyle 2\mu}
		\right\}
		\nonumber
		\\[0,2cm]
		&\stackrel{\mathscr{D}}{\longrightarrow}
		N\left(0,
		\left[
		{1\over y^2} \, \xi^{(1)}_g
		-
		{x\over y^3} \, \xi^{(1,2)}_g
		+
		{x^2\over 4y^4} \, \xi^{(2)}
		\right]
		\bigg\vert_{x=\mathbb{E}\left[ g(X_1,X_2)\right],y=\mu}
		\right). 
		\label{1-conv}
	\end{align}

	\smallskip
	Similarly, under the conditions  $\mathbb{E}\left[h^2(X_1,X_2)\right]<\infty$, {\color{black} where $h(X_1,X_2)$ is  defined in \eqref{def-h-new}}, and
	$\mathbb{E}[X_1^2]<\infty$, we obtain that,  as $n\to\infty$,
	\begin{align}\label{2-conv}
	\sqrt{n}(\widehat{H_{q}}-H_q)
	\stackrel{\mathscr{D}}{\longrightarrow}
	N\left(0,
	\left[
	{1\over y^2} \, \xi^{(1)}_h
	-
	{x\over y^3} \, \xi^{(1,2)}_h
	+
	{x^2\over 4y^4} \, \xi^{(2)}
	\right]
	\bigg\vert_{x=\mathbb{E}\left[ h(X_1,X_2)\right],y=\mu}
	\right),
	\end{align}
 where $\xi^{(1)}_h, \xi^{(1,2)}_h$ and $\xi^{(2)}$ are constructed in analogy with Equation \eqref{def-xi}.

	\begin{remark}
The significance of the convergence results \eqref{1-conv} and \eqref{2-conv} lies in its applicability to constructing confidence intervals and performing hypothesis tests in the context of large samples.
	\end{remark}

	\begin{remark}
Since $g(X_1,X_2)$ and $h(X_1,X_2)$ are non negative random variables, and $g(X_1,X_2)\leqslant\max\{X_1,X_2\}$ and $h(X_1,X_2)\leqslant 2\max\{X_1,X_2\}$  (see Remark \ref{rem-initial}), we have $\mathbb{E}\left[g^2(X_1,X_2)\right]\leqslant \max^2\{X_1,X_2\}\leqslant (X_1+X_2)^2\leqslant 2(X_1^2+X_2^2)$ and $\mathbb{E}\left[h^2(X_1,X_2)\right]\leqslant 4\max^2\{X_1,X_2\}\leqslant 4(X_1+X_2)^2\leqslant 8(X_1^2+X_2^2)$. Then, a sufficient condition for $\mathbb{E}\left[g^2(X_1,X_2)\right]<\infty$ and $\mathbb{E}\left[h^2(X_1,X_2)\right]<\infty$ is that both random variables, $X_1$ and $X_2$, have finite second-order moments.
Moreover, assuming that $X_1$ possess finite second moments, Lyapunov’s inequality ensures that condition $\mathbb{E}\left[X_1^2\right]<\infty$ is fulfilled.
	\end{remark}

\section{Simulation study}\label{Simulation study}

This section presents a Monte Carlo simulation to evaluate the finite-sample performance of the estimators defined in equations~(3) and~(4), namely $\widehat{G}_p$ and $\widehat{H}_q$. The objective is to assess their mean absolute relative error (MARE) and root mean squared error (RMSE) under different sample sizes and values of the parameter $p = q$. The MARE and RMSE of an estimator $\widehat{E}$ for a parameter $E$ were computed as follows:
$$
\widehat{\text{MARE}}(\widehat{E}) = \frac{1}{N_{\text{sim}}}  \sum_{k=1}^{N_{\text{sim}}}\left | \frac{\widehat{E}^{(k)} - E_{\text{true}}}{E_{\text{true}}}\right|,
$$
and
$$
\widehat{\text{RMSE}}(\widehat{E}) = \sqrt{ \frac{1}{N_{\text{sim}}} \sum_{k=1}^{N_{\text{sim}}} \big(\widehat{E}^{(k)} - E_{\text{true}} \big)^2 },
$$
where $E_{\text{true}} \in \{ G_p, H_q \}$ denotes the corresponding true value of the inequality measure, and $N_{\text{sim}}$ is the number of Monte Carlo replications.

The data were generated from a gamma distribution with shape parameter $\alpha = 1.5$ and scale parameter $\theta = 1$, so that the mean is $\mu = \alpha\theta = 1.5$. The simulation considered the following sample sizes: $n \in \{30, 50, 100, 200, 500\}$, and values of $p = q \in \{1.1, 2, 5, 10, 50\}$. For each combination of $(n, p)$, we generated $500$ independent samples. The true values $G_p$ and $H_q$, defined in Equations~\eqref{Gp} and~\eqref{Hq}, were approximated via large-sample Monte Carlo averages with $10^6$ observations.

\begin{figure}[!ht]
    \centering
    \includegraphics[width=0.6\textwidth]{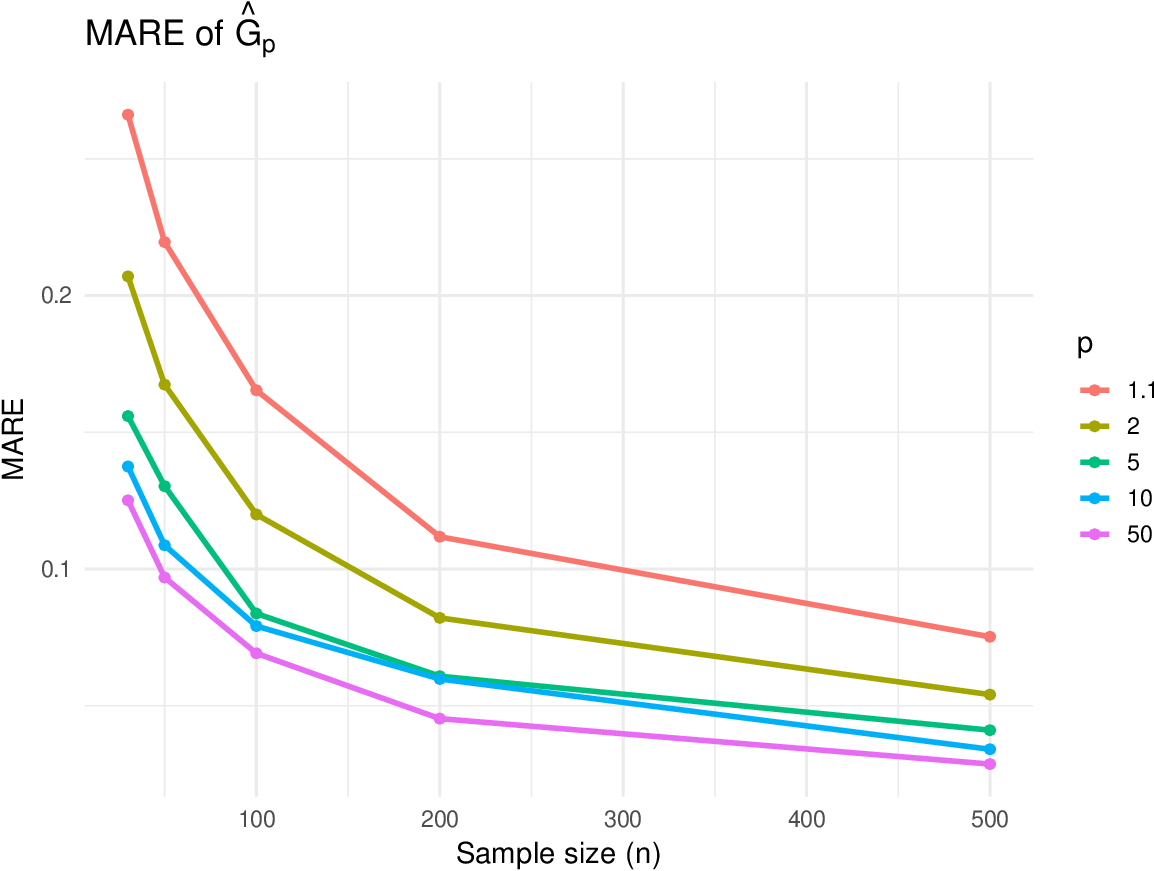}
    \caption{MARE of $\widehat{G}_p$ for varying sample sizes and values of $p$.}
    \label{fig:biasGp}
\end{figure}

\begin{figure}[!ht]
    \centering
    \includegraphics[width=0.6\textwidth]{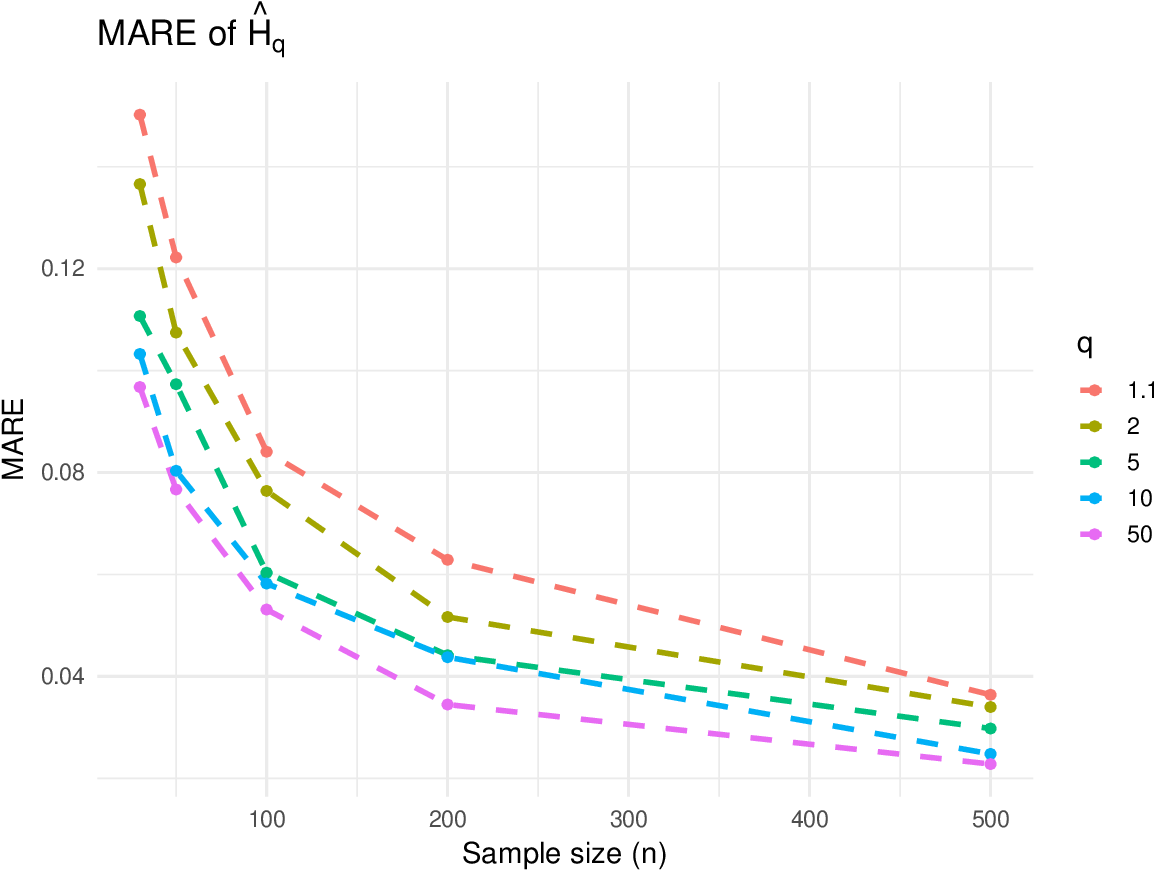}
    \caption{MARE of $\widehat{H}_q$ for varying sample sizes and values of $q$.}
    \label{fig:biasHq}
\end{figure}

\begin{figure}[!ht]
    \centering
    \includegraphics[width=0.6\textwidth]{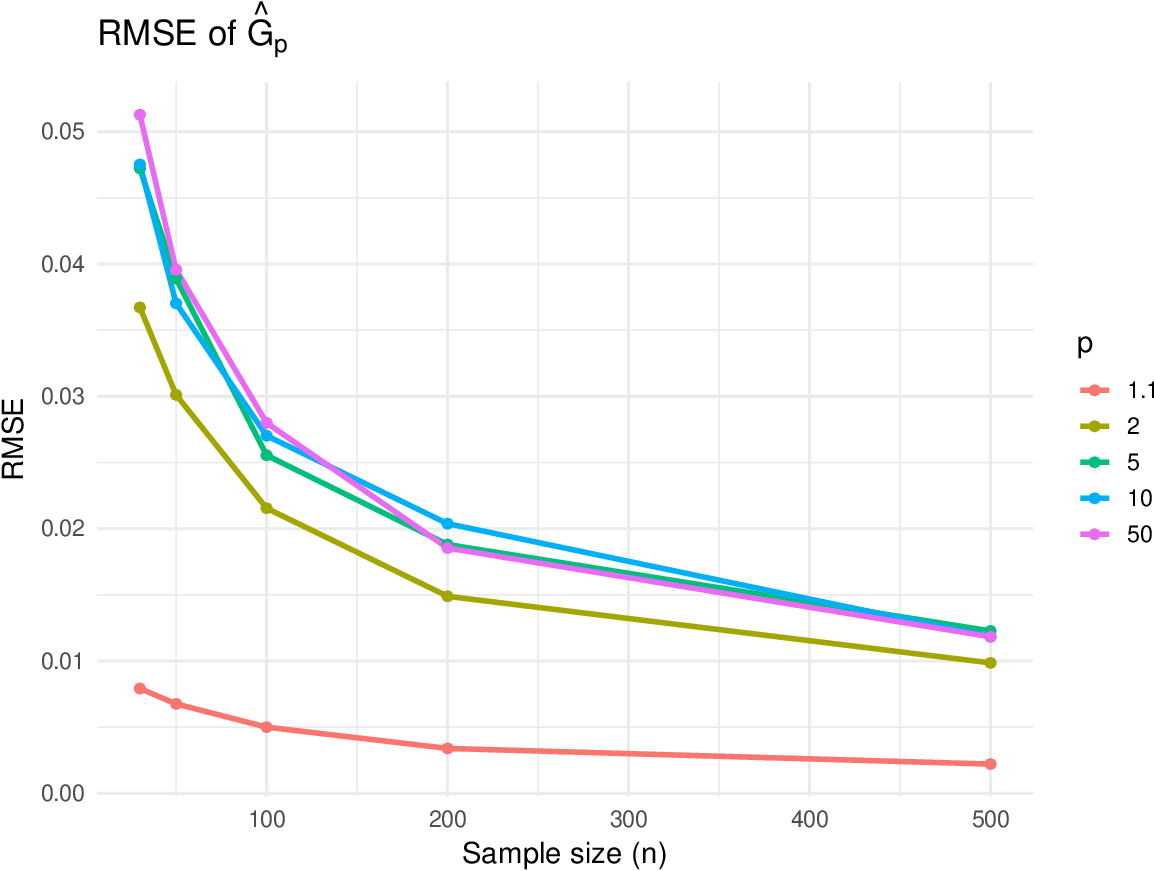}
    \caption{RMSE of $\widehat{G}_p$ for varying sample sizes and values of $p$.}
    \label{fig:rmseGp}
\end{figure}

\begin{figure}[!ht]
    \centering
    \includegraphics[width=0.6\textwidth]{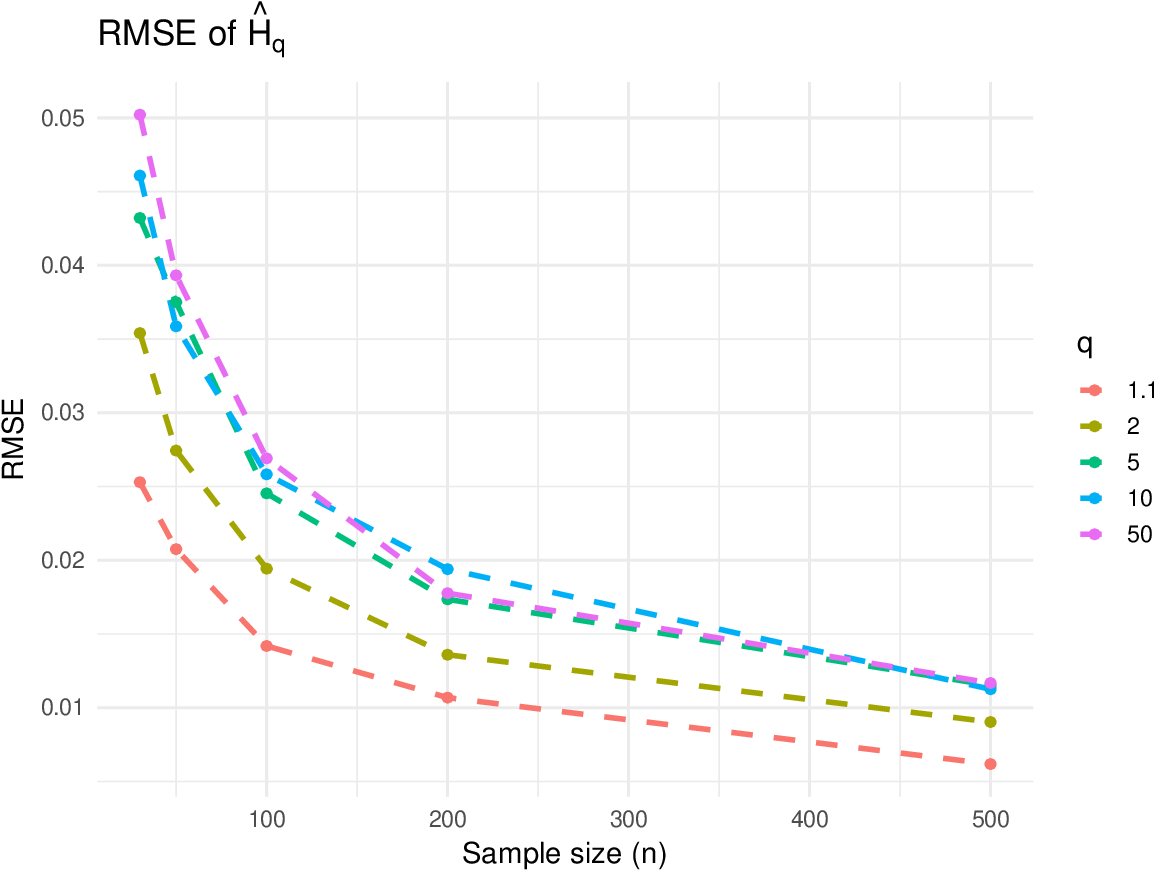}
    \caption{RMSE of $\widehat{H}_q$ for varying sample sizes and values of $q$.}
    \label{fig:rmseHq}
\end{figure}

The Monte Carlo simulation results are presented in Figures~\ref{fig:biasGp}--\ref{fig:rmseHq}. Particularly, Figures~\ref{fig:biasGp} and~\ref{fig:biasHq} display the MARE of the estimators $\widehat{G}_p$ and $\widehat{H}_q$, respectively, as functions of the sample size $n$, for different values of $p$, whereas Figures~\ref{fig:rmseGp} and~\ref{fig:rmseHq} show the RMSE behavior for the same configurations. From these figures, we observe that, as expected, the MARE and RMSE tend to decrease as the sample size $n$ increases. We also observe that larger values of $p$ tend to yield smaller (larger) MARE (RMSE). {\color{black} It is relevant to note that larger values of $p$ (or $q$) drive $G_p$ (or $H_q$) closer to the classical Gini coefficient, so the smaller MARE simply reflects that the estimator approximates a target that is numerically close to $G$ and inherits its favorable estimation properties; it does not imply that the Gini is always the preferred measure. Based on the simulation evidence, the asymptotic normal approximation established in Section~\ref{Asymptotic distribution} appears adequate for $n\geq 100$ across the parameter configurations considered. Furthermore, pilot studies with log-normal and Weibull distributions, with parameter values chosen to match the mean and coefficient of variation of the gamma scenario, yield qualitatively identical conclusions: both MARE and RMSE decrease monotonically with $n$ and the ordering of the estimators across values of $p$ (or $q$) is preserved.}

\section{Application}
\label{sec:application}

In this section, we illustrate the proposed income inequality measures using a data set on gross domestic product (GDP) per capita for all countries and territories in the Americas in 2023.  The raw data (in international dollars at 2021 prices) were downloaded from Our World in Data {\url{https://ourworldindata.org/grapher/gdp-per-capita-worldbank}} and converted into units of USD$\times10^3$. Our final sample comprises 34 countries, spanning from low‐ to high‐income contexts. We assume a gamma distribution for these data.

\begin{figure}[!ht]
\centering
\includegraphics[width=0.7\textwidth]{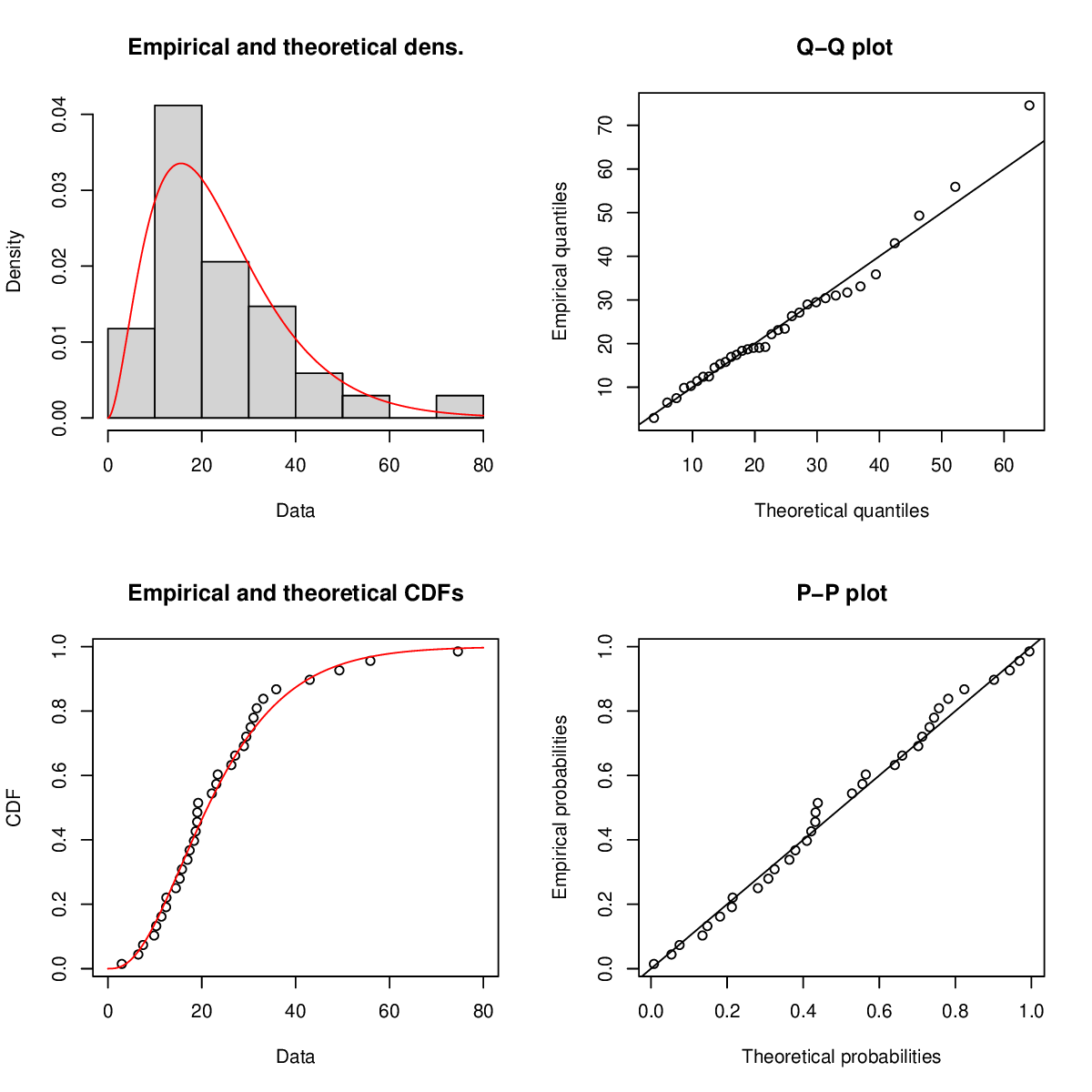}
\caption{Diagnostic plots for the gamma distribution fitted to GDP per capita data.}
\label{fig:diagnostic_plot}
\end{figure}

\begin{figure}[!ht]
    \centering
    \includegraphics[width=0.75\textwidth]{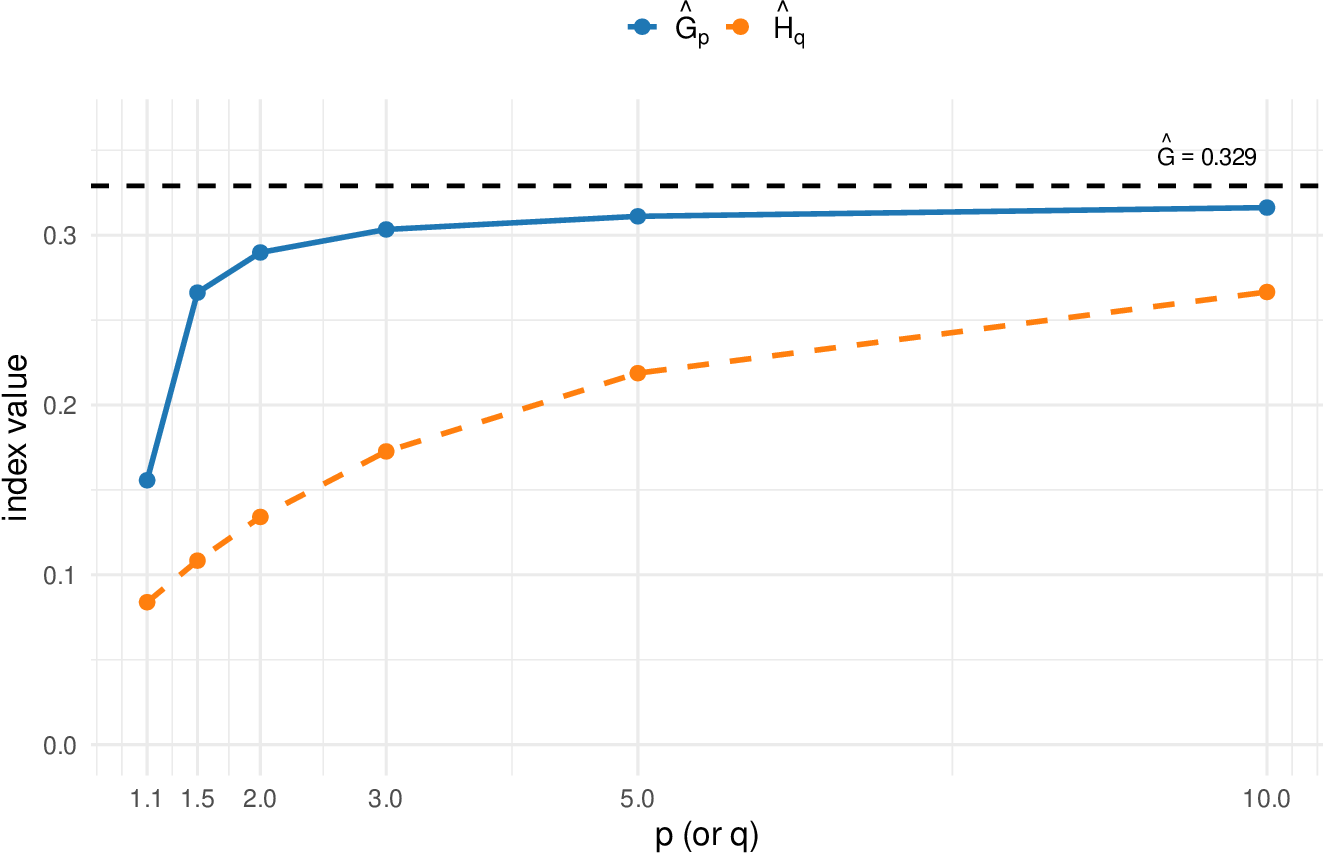}
    \caption{Estimated values of \(\widehat G_p\) and \(\widehat H_q\) for parameters \(p,q\in\{1.1,1.5,2,3,5,10\}\).  The dashed line indicates the classical Gini coefficient \(\widehat G\).}
    \label{fig:ineq_estimators}
\end{figure}

Figure~\ref{fig:diagnostic_plot} displays diagnostic plots based on the gamma distribution. From this figure, we observe that the gamma model can be a good choice for these data.  To further evaluate the adequacy of the gamma model, we performed two goodness‐of‐fit tests: the Kolmogorov–Smirnov (KS) test and the Cramér–von Mises (CvM) test. The respective $p$-values are 0.9151 and 0.9797, indicating no evidence to reject the gamma model. Hence, these results provide strong support for the adequacy of the gamma distribution assumption. {\color{black} Validating the gamma fit is important because the kernels $T(p)$ and $K(q)$, and consequently the theoretical behavior of $G_p$ and $H_q$ as functions of $p$ and $q$, depend on the underlying distribution; confirming that the gamma model is adequate allows us to connect the empirical estimates in Figure~\ref{fig:ineq_estimators} with the theoretical curves discussed in Section~\ref{Income inequality estimators}. The maximum likelihood estimates are: shape $\hat\alpha \approx 2.58$ and rate $\hat\lambda \approx 0.108$ (equivalently, scale $\hat\theta \approx 9.27$).}

We estimate the classical Gini coefficient $G $ and the generalized indices $G_p$ and $H_q$, for $p,q \in \{1.1,\,1.5,\,2,\,3,\,5,\,10\}$. Table~\ref{tab:estimates} reports the estimates for several values of $p\, (q)$. Figure~\ref{fig:ineq_estimators} displays the curves \(G_p\) and \(H_q\) as functions of $p$ and $q$, respectively, with the classical Gini \(\widehat G\) shown as a dashed horizontal line ($\widehat G = 0.329$). We note that, as the parameter $p$ ($q$) increases, both \(\widehat G_p\) and \(\widehat H_q\) increase, reflecting greater emphasis on the largest pairwise gaps in GDP per capita. {\color{black} From the profile of estimates in Figure~\ref{fig:ineq_estimators}, we observe that $\widehat{G}_p$ rises steeply from $p=1.1$ to $p=2$ and then levels off toward the Gini, whereas $\widehat{H}_q$ rises more gradually across the entire range. This indicates that extreme pairwise disparities are a notable driver of GDP inequality in the Americas, and that $G_p$ with moderate $p$ already captures most of this effect.

In order to guide practitioners in the selection of $p$ or $q$, we recommend the following approach. First, plot $\widehat{G}_p$ (or $\widehat{H}_q$) as a function of $p$ (or $q$) and examine the sensitivity profile. If the profile is relatively flat, inequality is spread uniformly across pairwise gaps and the classical Gini is an adequate summary; if the profile rises steeply, extreme disparities dominate and large $p$ (or $q$) is appropriate to emphasize them, while small $p$ (or $q$) down-weights extreme gaps and measures typical pairwise inequality. Second, if a specific normative framework prescribes a level of inequality aversion, $p$ (or $q$) can be calibrated to reflect the desired sensitivity, using the monotonicity property established in Proposition~\ref{Monotonicity} as a guide.

It is relevant to note that $H_q$ is ratio-scale invariant whereas $G_p$ is not; consequently, $H_q$ is the preferred choice when the measurement unit of income should not affect the inequality assessment, while $G_p$ may offer additional flexibility in settings where the scale of income is itself informative.

For the GDP data, the asymptotic normality results in Section~\ref{Asymptotic distribution} allow construction of approximate confidence intervals: a $100(1-\alpha)\%$ interval for $G_p$ takes the form $\widehat{G}_p\pm z_{\alpha/2}\,\widehat\sigma_{G_p}/\sqrt{n}$, where $\widehat\sigma_{G_p}$ is a consistent plug-in estimate of the asymptotic standard deviation from equation~\eqref{1-conv}; an analogous interval holds for $H_q$. Given $n=34$, bootstrap confidence intervals provide a reliable finite-sample alternative.

For $H_q$, the influence of $q$ is more nuanced: small values of $q$ yield a power-mean difference $K(q)$ close to the arithmetic-harmonic mean difference, distributing weights relatively uniformly across pairs; as $q$ increases, the power means approach the pair maximum and minimum respectively, and $H_q$ assigns progressively greater weight to the most extreme pairwise gaps.}

\begin{table}[!ht]
  \centering
  \caption{Estimated values of \(\widehat G_p\) and \(\widehat H_q\) for several values of \(p\) (or \(q\)), based on GDP data.}
  \label{tab:estimates}
  \begin{tabular}{rcc}
    \toprule
     $p$ ($q$) & \(\widehat G_p\) & \(\widehat H_q\) \\
    \midrule
    1.1  & 0.1557 & 0.0839 \\
    1.5  & 0.2662 & 0.1084 \\
    2.0  & 0.2898 & 0.1341 \\
    3.0  & 0.3034 & 0.1727 \\
    5.0  & 0.3111 & 0.2188 \\
   10.0  & 0.3163 & 0.2666 \\
    \bottomrule
  \end{tabular}
\end{table}

\section{Concluding remarks}\label{Concluding remarks}

In this paper we have introduced two flexible inequality measures, \(G_p\) and \(H_q\), which generalize the classical Gini coefficient. By deriving closed‐form U‐statistic estimators for each index, we established strong consistency and asymptotic normality under mild moment conditions. We carried out a Monte Carlo simulation to evaluate the performance of the proposed estimators \(\widehat G_p\) and \(\widehat H_q\), and the results have suggested that both the mean absolute relative error and root mean squared error tend to decrease as the sample size increases, as expected. An empirical illustration using GDP per capita in the Americas demonstrated how practitioners can select \(p\) or \(q\) to regulate the influence of disparities between observations. From a policy perspective, these indices may help to improve inequality analysis, as different weights can be attributed to disparities between observations.

\paragraph*{Acknowledgements}
The research was supported in part by CNPq and CAPES grants from the Brazilian government.

	\paragraph*{Disclosure statement}
	There are no conflicts of interest to disclose.
%
%
%
%
%
%
%
%

%
%
%
\end{document}